\documentclass[reprint,superscriptaddress, nofootinbib,longbibliography]{revtex4-2}
\pdfoutput=1
\usepackage{graphicx}
\usepackage{amssymb}
\usepackage[linktoc=none]{hyperref}
\usepackage[dvipsnames]{xcolor}
\usepackage{dcolumn}
\usepackage[english]{babel}
\usepackage[utf8x]{inputenc}
\usepackage[T1]{fontenc}
\usepackage{eqnarray,amsmath,physics,bbold}
\usepackage{ dsfont }
\usepackage{appendix}
\usepackage{xcolor}
\usepackage{breqn}

\usepackage[normalem]{ulem}

\usepackage{hyperref}
    \hypersetup{
    colorlinks   = true,    % Colours links instead of ugly boxes
    urlcolor     = blue,    % Colour for external hyperlinks
    linkcolor    = blue,    % Colour of internal links
    citecolor    = red      % Colour of citations
    }

\DeclareMathAlphabet{\mathpzc}{OT1}{pzc}{m}{it}

\newcommand{\INFN}{INFN - Sezione collegata di Salerno, Complesso Univ. Monte S. Angelo, I-80126 Napoli, Italy}

\newcommand{\UNISA}{Physics Department ``E.R. Caianiello'', Universit\`a degli studi di Salerno, Via Giovanni Paolo II, 132, I-84084 Fisciano (Sa), Italy}

\newcommand{\CNR}{CNR-SPIN, I-84084 Fisciano (Salerno), Italy, c/o Universit\`a di Salerno, I-84084 Fisciano (Salerno), Italy}

\newcommand{\IFW}{Institute for Theoretical Solid State Physics, IFW Dresden, Helmholtzstr. 20, 01069 Dresden, Germany}

\newcommand{\ctqmat}{W\"{u}rzburg-Dresden Cluster of Excellence ct.qmat, Helmholtzstrasse 20, 01069 Dresden, Germany}

\newcommand{\ARGENTINA}{Centro Atomico Bariloche, Instituto de Nanociencia y Nanotecnologia (CNEA-CONICET) and Instituto Balseiro, Av. Bustillo, 9500, Argentina}

\begin{document}
\title{
Non-linear anomalous Edelstein response at altermagnetic interfaces}

\author{Mattia Trama}
\email{mtrama@unisa.it}
\affiliation{\IFW}
\affiliation{\ctqmat}
\affiliation{\UNISA}

\author{Irene Gaiardoni}
\affiliation{\UNISA}

\author{Claudio Guarcello}
\affiliation{\UNISA}
\affiliation{\INFN}

\author{Jorge I. Facio}
\affiliation{\ARGENTINA}

\author{Alfonso Maiellaro}
\affiliation{\UNISA}
\affiliation{\CNR}

\author{Francesco Romeo}
\affiliation{\UNISA}
\affiliation{\INFN}

\author{Roberta Citro}
\email{rocitro@unisa.it}
\affiliation{\UNISA}
\affiliation{\CNR}
\affiliation{\INFN}

\author{Jeroen van den Brink}
\affiliation{\IFW}
\affiliation{\ctqmat}

\begin{abstract}
In altermagnets, time-reversal symmetry breaking  spin-polarizes electronic states, while total magnetization remains zero. In addition, at altermagnetic surfaces Rashba-spin orbit coupling is activated due to broken inversion symmetry, introducing a competing spin-momentum locking interaction. Here we show that their interplay leads to the formation of complex, chiral spin textures that offer novel, non-linear spin-to-charge conversion properties. Whereas altermagnetic order suppresses the canonical linear in-plane Rashba-Edelstein response, we establish the presence of an \textit{anomalous} transversal Edelstein effect for planar applied electric and magnetic field, or alternatively, an in-plane magnetization. 
Additionally, we predict a purely electric-field-driven non-linear out-of-plane magnetization. We compute the anomalous response within a general altermagnet $d$-wave model, with parameters extracted from the ab-initio electronic structure of an altermagnetic bilayer. 
Our results suggest altermagnetic surfaces as a promising platform for unconventional spintronic functionalities.
\end{abstract}

\maketitle

\section{Introduction}
\begin{figure*}[ht!!]
    \centering
    \includegraphics[width=0.98\textwidth]{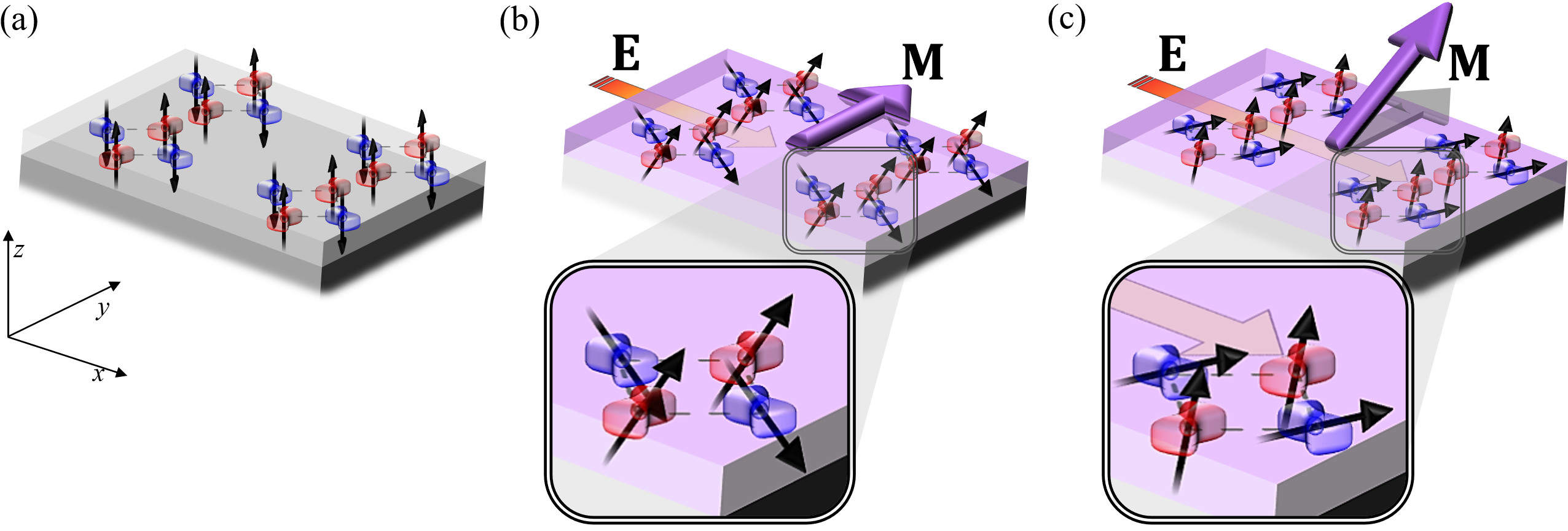}
    \caption{\textbf{Schematic representation of the transversal, non-linear altermagnetic EE}. (a) Out-of-plane spin arrangement in the absence of driving current with zero net magnetization. (b) An in-plane driving current tilts the out-of-plane spins resulting in an in-plane transverse magnetic field. (c) A stronger current is able to further tilt the spins resulting in a non-linear magnetization, which is also out-of-plane.}
    \label{fig:cartoon}
\end{figure*}
Altermagnets (AMs) are a class of magnetic materials with a total magnetization that vanishes by symmetry while at the same time  Kramers' degeneracy is lifted~\cite{smejkal2022beyond,mazin2021prediction,hayami2019momentum,smejkal2020crystal,naka2019spin,yuan2020giant,naka2021perovskite,guo2023spins,sato2024altermagnetic,smejkal2022anomalous,hayami2019momentum,ma2021multifunctional}. 
Unlike antiferromagnets, for which the vanishing of the magnetization is enforced by the combined action of time-reversal and translation/inversion symmetry, in AMs this involves an additional rotation or mirror operation. This symmetry operation causes the band structure to be spin-polarized, also in the absence of spin-orbit coupling (SOC)~\cite{smejkal2022beyond}. 
The presence of spin-polarized bands and Fermi surfaces makes AMs potentially appealing as spin-polarized current sources for antiferromagnetic spintronics~\cite{naka2019spin,bai2023efficient,gonzalez2021efficient,zelezny2017spin,bai2022observation,bose2022tilted,golub2025spinorientationelectriccurrent}.
On the basis of magnetic symmetries~\cite{smejkal2022beyond}, an extensive set of candidate materials exhibiting collinear altermagnetism have been identified from electronic structure calculations, both 3D~\cite{guo2023spins} and 2D~\cite{milivojevic2024interplay}. 
While experimentally the spin-splitting of electronic bands has been observed by photoemission spectroscopy~\cite{zeng2024observation,reimers2024direct,lu2024observation,jiang2024discovery,zhang2024crystal}, exploration of AM (magneto--) transport properties for antiferromagnetic spintronic purposes, which capitalizes on their strong exchange interactions~\cite{daldin2024antiferromagnetic},  has only just begun and has so far focused on harnessing their anomalous and spin Hall effects~\cite{li2024topological,sato2024altermagnetic,gonzalez2021efficient,feng2022an,gonzalez2023spontaneous,bai2022observation,reimers2024direct,hu2024spinhalledelsteineffects}. For instance it has recently been shown that altermagnetic spin-splitting is at the origin of a spin Hall effect in RuO$_2$~\cite{bai2023efficient}.
In this context, we identify natural possibilities for spin-to-charge conversion processes at AM interfaces that fundamentally go beyond the canonical linear Rashba-Edelstein effect~\cite{edelstein_effect, trama2022tunable,trama2024charge},
namely the generation of an in-plane magnetization in response to a planar electric field caused by antisymmetric spin-texture at the Fermi surface. Indeed, we show that this response is actually suppressed by AM.
%where 
%a surface electric current produces an in-plane magnetization that, as we will show, is actually suppressed by AM. 
Instead, the competition between Rashba SOC (RSOC) and AM triggers \textit{anomalous} non-linear Edelstein effects (EEs): we will show that a magnetization transverse to the interface is induced by an in-plane applied electric and magnetic field (or an in-plane magnetization) -- or alternatively as non-linear magnetic response to purely planar electric fields.
This spin-to-charge conversion in electric and magnetic field can be depicted by an in-plane current forcing the magnetic moments out of the plane, as schematically indicated in Fig.~\ref{fig:cartoon}. Also the purely electric non-linear case will be understood in such basic terms.

It should be noted that, while AM may be strictly defined in the limit of vanishing SOC, it still defines an essential energy scale, determining for instance the altermagnetic anomalous Hall response~\cite{sato2024altermagnetic} and electronic properties related to the topology of altermagnetic electronic structures~\cite{li2024topological}.
At interfaces and in 2D, the RSOC is induced by inversion-symmetry breaking and naturally produces a spin-momentum locking without net magnetization. Indeed, the coexistence of AM spin splitting and the RSOC has been pointed out~\cite{sun2023spin,amundsen2024rkky,chen2024helicitycontrolledspinhall}, but so far not investigated as a mechanism to achieve an anomalous Rashba-Edelstein response and non-linear spin-to-charge conversion. 

\begin{figure}[hb!!]
 \centering
 \includegraphics[width=0.49\textwidth]{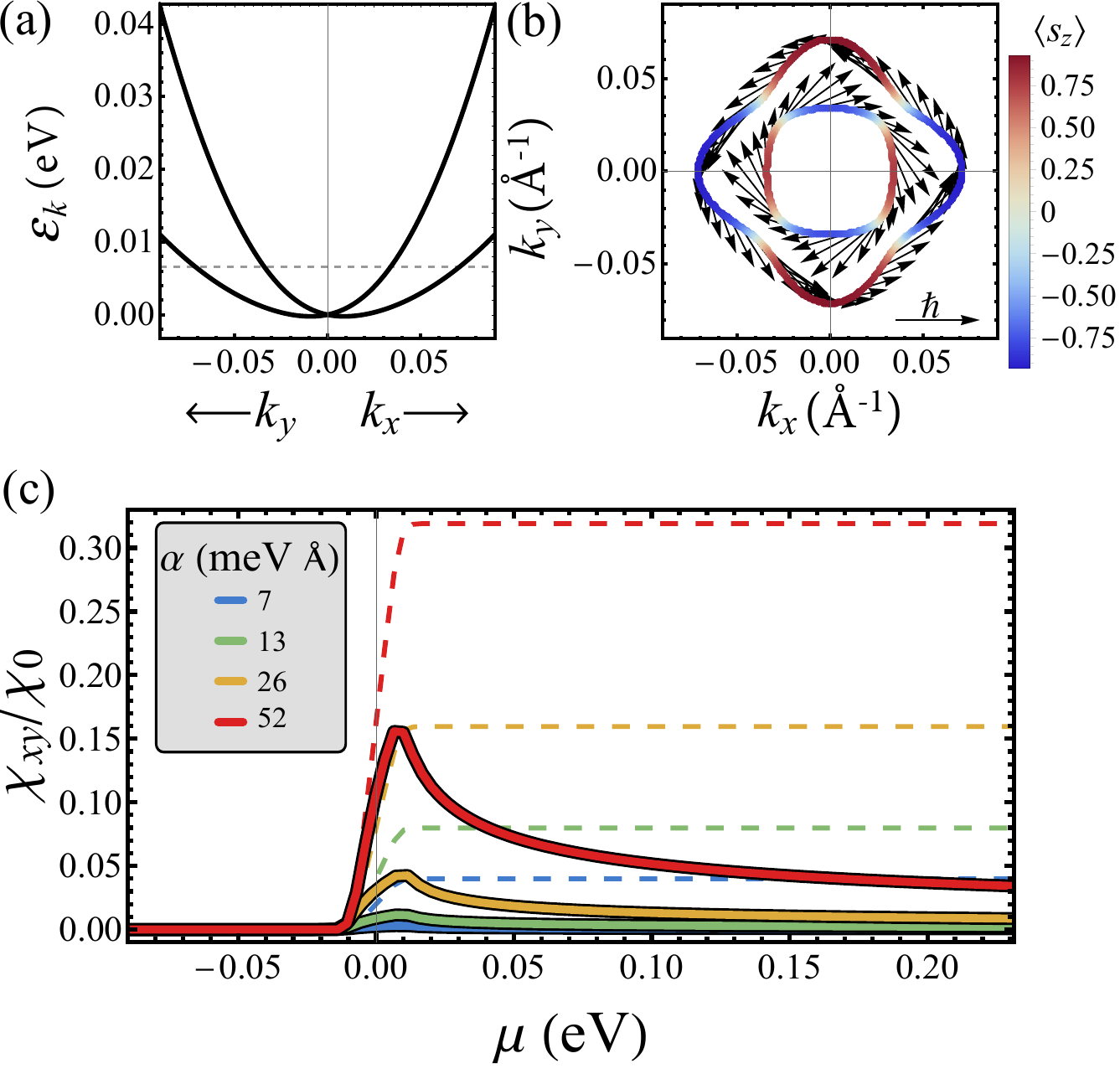}
 \caption{\textbf{Competing altermagnetic and Rashba spin-orbit interactions}. (a) Electronic dispersion resulting from the continuum Hamiltonian~\eqref{eq:ham_altermagnets_rashba} $\alpha=52$ meV \AA. (b) Fermi surface and spin structure for fixed chemical potential (dashed line in panel (a)). (c) linear Edelstein susceptibility $\chi_{xy}/\chi_0$ ($\chi_0=2.44\times10^{2}$ $\mu_B$~\AA~V$^{-1}$) for different $\alpha$. The dashed lines correspond to the pure RSOC case ($\gamma=0$). 
 }
 \label{fig:spin_density_rashba}
\end{figure}

The standard linear EE originating from RSOC locks the spin of electrons in-plane and orthogonal to their momentum, respecting spinful time-reversal symmetry. Then, an applied in-plane electric field results in an imbalance in the spin density at the Fermi surface and a net in-plane magnetization orthogonal to the induced current. Interestingly, in the presence of only RSOC, the second-order non-linear EE~\cite{baek2024nonlinear,vignale2016theory,xu2021light} vanishes by symmetry. However, it may be triggered by, for instance, introducing orbital degrees of freedom~\cite{baek2024nonlinear} or a time-dependent setting~\cite{vignale2016theory}. 
In contrast to RSOC, AM breaks time-reversal symmetry, so that in collinear AMs opposite momenta carry the same spin orientation. This renders the momentum-space magnetic textures generated by RSOC and AM fundamentally different: the two interactions compete.

\section{Altermagnetic Rashba model}
To explore the physical consequences of this competition, we start by considering the 2D effective continuum Hamiltonian 
\begin{equation}
\hat{H}= \frac{k^2}{2m} + \gamma (k_x^2 - k_y^2)\hat{\sigma}_z + \alpha \hat{z} \cdot (\vec \sigma \times \mathbf{k})-\mu,
\label{eq:ham_altermagnets_rashba}
\end{equation}
where $\hat{\sigma}_i$ is the Pauli matrix in the direction $\hat{i}$, $\alpha$ is the RSOC interaction, $\gamma$ is the altermagnetic $d$-wave coupling with symmetry $[C_2||C_{4z}]$~\cite{smejkal2022beyond}, $\mu$ is the chemical potential, and $\hbar$ is set to unity. The resulting bands and their spin-texture for given chemical potential are illustrated in Fig.~\ref{fig:spin_density_rashba}(a,b). Note that the chirality of the resulting non-coplanar spin texture is determined by $\mathrm{sgn}(\gamma\alpha)$.
To maintain contact with real materials, we choose the numerical values of the altermagnetic coupling $\gamma$ and the effective mass $m$ by explicitly comparing our $\bf k\cdot p$ model with the local band structure close to the $\Gamma$ point of a  RuO$_2$ bilayer. Details of ab-initio calculation and band-structure fitting are given in the Appendix~\ref{app:fit}. We use $m=0.152$~eV$^{-1}$~\AA$^{-2}$ and $\gamma=1.881$~eV~\AA$^{2}$. While the magnetic nature of RuO$_2$ is a debated question~\cite{lin2024observation, liu2024absence,jeong2024altermagnetic,kiefer2024crystal,wenzel2024fermi}, thin films of RuO$_2$~\cite{brahimi2024confinement,song2024spin,guod2024direct,kobayashi2024detection,Fedchenko_2024} have been characterized as $d$-wave altermagnets. Here, we consider this system just as a representative $d$-wave altermagnetic model , in which both the band filling and the effective RSOC can, in principle, be tuned via gating.
%
% \begin{figure}
%  \centering
%  \includegraphics[width=0.45\textwidth]{Figure_3.pdf}
%  \caption{\textbf{Bilayer of RuO$_2$ and its electronic structure density-functional calculations}. (a) Crystal structure. (b) Band structure along high symmetry points indicated in (c).}
%  \label{fig:band_strucure}
% \end{figure}

\subsection*{Linear Edelstein responses} The linear EE is defined through $M_j=\chi_{ij}E_i$, where $E_i$ is the electric field along the direction $\hat{i}$ and $\chi_{ij}$ is the Edelstein susceptibility. The spin accumulation $M_j$ can be obtained in the Boltzmann framework as
\begin{equation}
 M_j(\mu)=-\mu_b\sum_{\mathbf{k}, \nu}|e|\left(\vec{\Lambda}_{\mathbf{k}}^\nu \cdot \mathbf{E}\right) \delta\left[\varepsilon_{\mathbf{k}}^\nu-\mu\right](\nu s_j(\mathbf{k})),
 \label{eq:ed_sum}
\end{equation}
where we made $\mu$ explicit in the Dirac delta since we treat it as an external parameter, $\mu_b$ is the Bohr magneton, $\varepsilon_{\mathbf{k}}$ are the eigenvalues with band index $\nu=\pm$ that, for the Hamiltonian in Eq.~\eqref{eq:ham_altermagnets_rashba}, are given by
\begin{equation}
    \varepsilon_{\mathbf{k}}^\pm\!=\!\frac{k^2}{2m}\pm\sqrt{\alpha^2 (k_y^2+ k_x^2)+\gamma^2(k_x^2-k_y^2)^2}\!=\!\frac{k^2}{2m}\pm \delta\varepsilon,
    \label{eq:eigens}
\end{equation}
and $s_j$ is the $j-$component of the spin vector $\mathbf{s}=\left(\alpha k_y ,-\alpha k_x, \gamma(k_x^2-k_y^2)\right)/\delta\varepsilon$.
The mean free path is computed within the relaxation-time approximation, $\vec{\Lambda}_{\mathbf{k}}^{\nu}$= $\tau_{\mathbf{k}}^{\nu} \mathbf{v}_{\mathbf{k}}^{\nu}$, with  $\tau_{\mathbf{k}}^{\nu}$ and $\mathbf{v}_{\mathbf{k}}^{\nu}$= $\nabla_{\mathbf{k}} \varepsilon^\nu_{\mathbf{k}}$ being the transport lifetime and the group velocity, respectively.

Supposing, for simplicity, that $\tau^\nu_{\mathbf{k}}=\tau$, and taking $\mathbf{E}=E_0\hat{i}$, Eq.~\eqref{eq:ed_sum} can be written in terms of the susceptibility per unit cell, $\chi_{ij}$, as 
\begin{equation}
    \chi_{ij}=-a\chi_0 \sum_{\nu=\pm}\int d^2\mathbf{k}\; \nu s_j(\mathbf{k})\delta(\varepsilon_{\mathbf{k}}^\nu-\mu)v^\nu_i(\mathbf{k}),
    \label{eq:edel_suscept}
\end{equation}
where $\chi_0=\frac{\tau |e| \mu_b S_{\text{cell}}}{4\pi^2 a}$, with $a$ being the lattice parameter, $ S_{\text{cell}}$ the area of the unit cell, and $\tau=10^{-12}$~s the typical order of magnitude of oxides~\cite{trama2022tunable}.

The conventional EE is governed by $\chi_{xy}$ and RSOC alone is sufficient to induce it, as shown for our continuum Hamiltonian in Fig~\ref{fig:spin_density_rashba}(c). 
The symmetry group of the Hamiltonian $[C_2||C_{4z}]$ with broken inversion symmetry enforces $\chi_{xx}=\chi_{yy}=0$, $\chi_{xy}=-\chi_{yx}$, and $\chi_{iz}=0$ for $i=x,y$.
%This can be confirmed by symmetry considerations: the group of symmetry of Hamiltonian~\eqref{eq:ham_altermagnets_rashba} is $\{\mathcal{T}C_{4z},\mathcal{T}\mathcal{M}_{x}\}$, where $\mathcal{T}$ is the time-reversal operator, and $\mathcal{M}_{x}$ is the mirror symmetry with respect to $yz$ plane. Since $\chi_{ij}\propto\frac{\partial s_j}{\partial k_i}$, it is true that $\chi_{ii}=0$, $\chi_{xy}=-\chi_{yx}$ and $\chi_{iz}=0$ for $i=x,y$.}
Without AM and in-plane magnetic field ($\gamma=0$), the linear Edelstein susceptibility $\chi_{xy}$ scales with the RSOC $\alpha$, and quickly saturates for large chemical potential $\mu$, see dashed lines in Fig~\ref{fig:spin_density_rashba}(c). Here, we vary $\alpha$ in a range around $0.05$ eV \AA~which is a realistic range of values for oxides~\cite{lesne2016highly,trama2022gate,trama2021strain,zhai2024large}, and may be experimentally achieved by, e.g., a tunable top-gate or interfacial electric field~\cite{caviglia2010}.
Clearly, a finite altermagnetic coupling, $\gamma$, significantly suppresses the EE, particularly at higher $\mu$, and this for any value of $\gamma$ leads to a pronounced peak in $\chi_{xy}$ in the vicinity of the band bottom. Thus, as we expected on the basis of symmetry considerations, AM and the conventional EE strongly compete.

\subsection*{Definition of non-linear Edelstein responses} 
Clearly, the RSOC, with its in-plane spin-momentum locking, cannot alone produce an out-of-plane, transverse Edelstein response, i.e., $\chi_{xz}$ vanishes identically. 
Even if the $\gamma$ term tends to align spins out-of-plane, RSOC in the presence of altermagnetism  is not sufficient to induce a transversal response. Such an anomalous transversal response actually requires altermagnetism and higher orders in the fields, as we will detail in the following. One may distinguish two types of non-linear (second order) responses, one involving an in-plane electric and magnetic field and one involving only in-plane electric fields to second order.
In the following, we focus solely on the extrinsic contribution arising from impurity scattering. For such a model, the intrinsic interband Edelstein contribution, originating from the off-diagonal terms~\cite{li2015intraband}, vanishes by symmetry (see details in Appendix~\ref{app:intrinsic}). Other intrinsic effects, such as those related to quantum geometry~\cite{xiao2022intrinsic} or deviations from the rigid-band approximation~\cite{jia2024equivalence,kaplan2024unification}, could in principle be considered, but they are expected to be more relevant for complex multiband structures. Here, we highlight the standard contribution, which should also dominate for thin films on a substrate, where the constant-scattering-time approximation is valid~\cite{Johansson2021spin}.

The case of combined electric and magnetic fields is most easily considered by introducing a Zeeman term in Eq.~\eqref{eq:ham_altermagnets_rashba}, orthogonal to the applied electric field $E_x$, of the form $H_{h}=h_y\sigma_y$ and then using Eq.~\eqref{eq:edel_suscept}. 
The $H_{h}$ term corresponds to an effective homogeneous spin splitting, which may arise from a Zeeman field or from weak ferromagnetism induced by Dzyaloshinskii–Moriya (DM) interactions, generally allowed by AM symmetries~\cite{wang2022magneto,milivojevic2024interplay,cheong2024altermagnetism,autieri2024staggered}.
The presence of the \textit{in-plane} magnetic field spontaneously breaks the $\mathcal{T}C_{4z}$ symmetry, allowing out-of-plane Edelstein response.
This response is nonlinear as it depends both linearly on electric field {\it and} magnetic coupling and vanishes when either one is zero (see Appendix~\ref{app:spin_text_an}). Incorporating  $\mathbf{h}$ directly into the Hamiltonian offers the advantage that it is easy to go beyond its linear response.  

To assess the non-linear EE involving only electric fields, it is more convenient to introduce the second order Edelstein susceptibility of the form $M_j=\Tilde{\chi}_{iij}E_i^2$, with
\begin{equation}
    \Tilde{\chi}_{iij}=-\Tilde{\chi}_{0} \sum_{\nu=\pm}\int d^2 \mathbf{k} \;  \nu \frac{\partial  s_j(\mathbf{k})}{\partial k_i}  \delta(\varepsilon^\nu_{\mathbf{k}}-\mu) v_i^\nu(\mathbf{k}),
    \label{eq:tilde_chi_eq}
\end{equation}
where $\tilde{\chi}_0=\frac{\mu_B e^2 \tau^2}{4\pi^2} S_{\text{cell}}$ (see Appendix~\ref{app:analytical} for the full derivation). 
The $\mathcal{T} C_{4z}$ symmetry does not rule out an out-of-plane response and enforces $\tilde{\chi}_{xxz}=-\tilde{\chi}_{yyz}$, (while it easily excludes an in-plane spin response $\tilde{\chi}_{iij}=0$ for $i,j=x,y$).
\begin{figure}[t!!]
 \centering
  \includegraphics[width=0.49\textwidth]{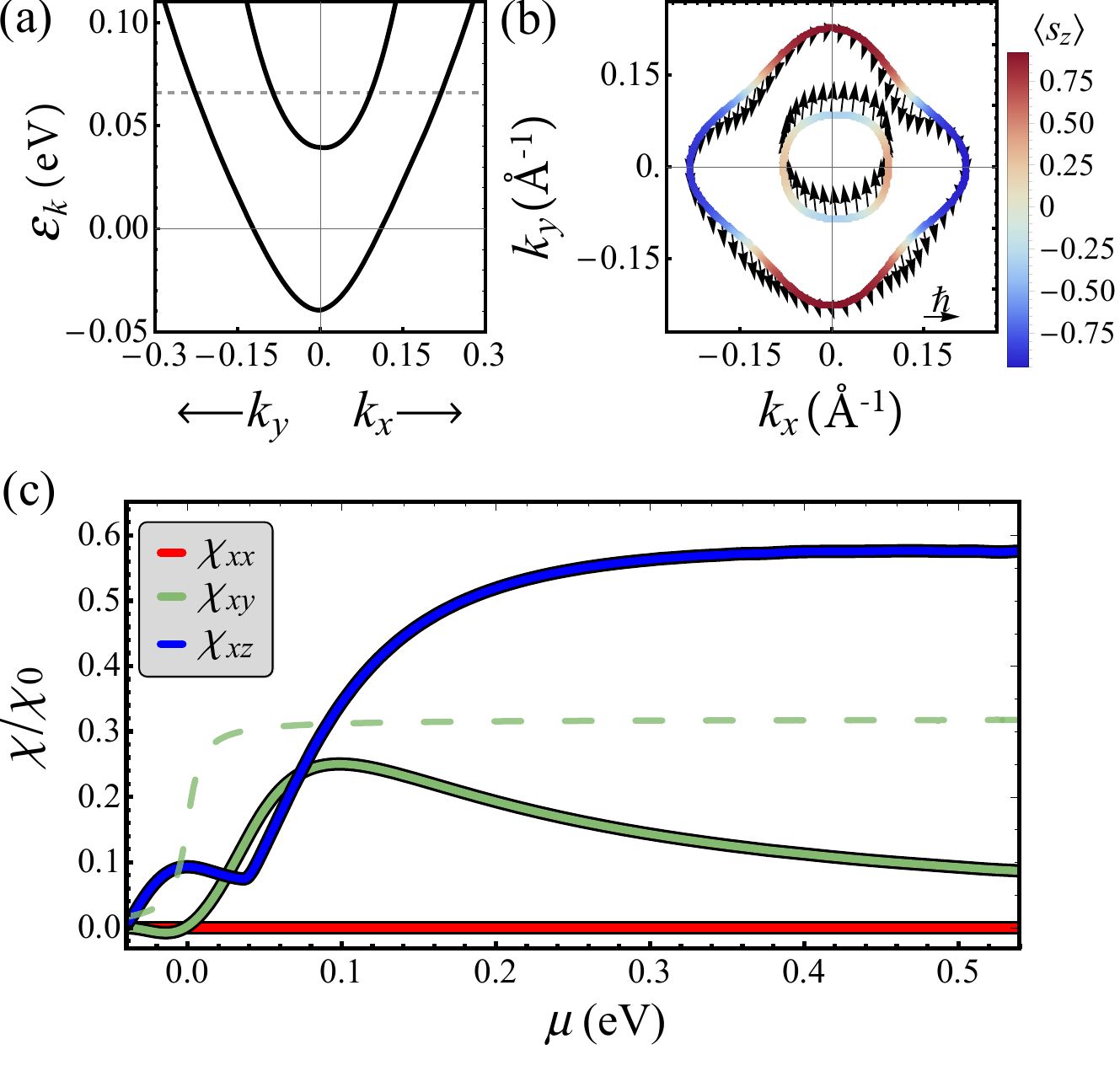}
 \caption{\textbf{Anomalous non-linear EE due to a combined planar magnetic and electric field}. (a) Electronic dispersion from Hamiltonian~\eqref{eq:ham_altermagnets_rashba} with $\alpha=52$ meV \AA~and $h_y=40$ meV. (b) Fermi surface and spin structure for fixed chemical potential, see the dashed line in (a). 
 (c) Edelstein response $\chi_{xi}/\chi_0$ ($\chi_0=2.44\times10^{2}$ $\mu_B$~\AA~V$^{-1}$) for the directions $i={x,y,z}$. The green dashed line corresponds to the pure RSOC case ($\gamma=h_y=0$).
 }
 \label{fig:Zeeman_edelstein}
\end{figure}

\section{Calculation of non-linear Edelstein responses} 
Having established that, due to the interplay of RSOC with AM symmetry allows anomalous non-linear EEs, we now determine them  
in an experimentally relevant parameter range.
For the effective in-plane field $h_y$ resulting from weak ferromagnetism, we choose $h_y=40$ meV as benchmark value~\cite{milivojevic2024interplay}. 
The first effect of the in-plane Zeeman coupling is to lift the degeneracy at $\Gamma$, as shown in Fig.~\ref{fig:Zeeman_edelstein}(a), in contrast with the simple Rashba system, for which the in-plane magnetic field can only move the degeneracy point. Moreover, as mentioned above, it breaks the symmetry of the system both in the shape of the Fermi surface and in the chiral structure of the spin, as one can see in Fig.~\ref{fig:Zeeman_edelstein}(b).
The resulting Edelstein susceptibility tensor $\chi$ is shown in Fig.~\ref{fig:Zeeman_edelstein}(c). While $\chi_{xx}$ vanishes, $\chi_{xy}$ depends non-monotonically on chemical potential and reaches its maximum after occupying the first band. Compared to the case $h_y=0$, $\chi_{xy}$ grows slower to its maximum, but reaches a value similar to the situation with pure RSOC. Moreover, the in-plane magnetic field is able to induce a substantial out-of-plane anomalous transversal response $\chi_{xz}$ due to the large altermagnetic coupling, with the transversal component reaching larger values than the in-plane response, even larger than the canonical linear RSOC situation.

The non-monotonic behaviour of $\chi_{xz}$ as a function of $\mu$ derives from the competition and complex interplay of the different interactions. At low filling, even an infinitesimal $h_y$ breaks the $\mathcal{T}C_{4z}$ symmetry, leading to a characteristic threshold behavior in the response. Increasing $\mu$, the out-of-plane response is first slightly suppressed and later saturates to a constant value in the altermagnetic-dominated regime. Crucially, all three ingredients are required to exhibit the out-of-plane response: the altermagnetic interaction enforces an out-of-plane magnetization, the RSOC induces the spin-momentum locking, and the weak ferromagnetism breaks its otherwise symmetric pattern required for this non-linear Edelstein response.

\begin{figure}[t!!]
 \centering
  \includegraphics[width=0.4\textwidth]{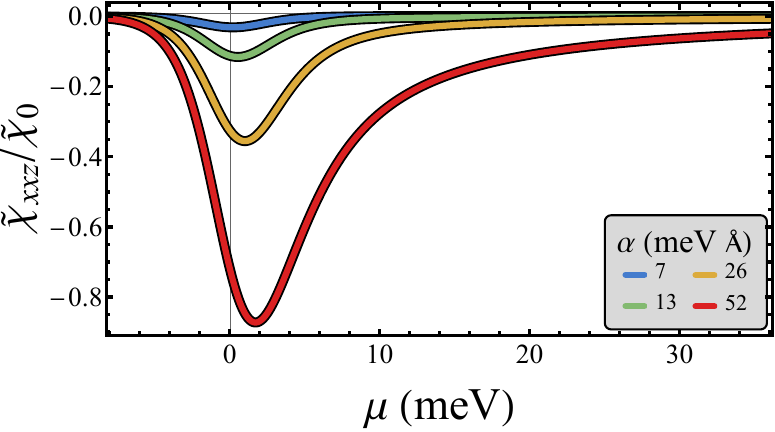}
  \caption{\textbf{Second-order electric Edelstein response}. $\tilde{\chi}_{xxz}/\tilde{\chi}_0$ (with $\tilde{\chi}_0=2.36 \times 10^{6} \,\text{\AA}^2\text{V}^{-2}$) versus $\mu$ for different values of $\alpha$.
 }
 \label{fig:second_chi_tilde}
\end{figure}

On the other hand, without inducing any modification to the energy spectrum, i.e., the electronic bands and the spin structure correspond to the one shown in Fig.~\ref{fig:spin_density_rashba}, an out-of-plane, second-order EE is obtained. The computed Edelstein susceptibilities from Eq.~\eqref{eq:tilde_chi_eq}, for different values of the RSOC, are shown in Fig.~\ref{fig:second_chi_tilde}.
One observes a signature peak right above the band crossing at $\varepsilon=0$, which decreases with the increasing of $\alpha$, and then a steady decrease of $\tilde{\chi}_{xxz}$ for larger $\mu$. This anomalous response is characteristic of altermagnetic systems with RSOC and vanishes for either $\alpha=0$ or $\gamma=0$ (see Appendix~\ref{app:analytical} for details). Therefore, this non-monotonic behaviour allows to individuate the location of band crossing due to RSOC and the strength of the coupling in altermagnetic surfaces.
A detailed analysis is presented in Appendix~\ref{app:parameters}.
When generalized to finite frequencies, the non-linear susceptibility $\Tilde{\chi}_{xxz}$ induces second-harmonic generation. This, in principle, offers an experimentally attractive route to detect such a non-linear response. In our case, the rectifying and doubled frequency terms $\expval{\tilde{\chi}_{xxz}}=\text{Re}[\expval{\tilde{\chi}_{xxz}^0}+\expval{\tilde{\chi}_{xxz}^{2\omega}}e^{2i\omega t}]$ appear, which are related to Eq.~\eqref{eq:tilde_chi_eq} as $\expval{\tilde{\chi}_{xxz}^0}=\tilde{\chi}_{xxz}/{2(1+i\omega\tau)
}$ and $\expval{\tilde{\chi}_{xxz}^{2\omega}}=\tilde{\chi}_{xxz}/{2(1+i\omega\tau)(1+2i\omega\tau)}$.

\section{Summary and outlook} 
Using a continuum description for AMs based on  
ab-initio electronic band structure calculations, we have shown that the interplay of altermagnetic coupling with an interfacial RSOC gives rise to the appearance of out-of-plane spin polarizations throughout the Brillouin zone via non-linear EEs. 
While the in-plane linear response can be strongly suppressed with the band filling, the non-linear Edelstein susceptibilities exhibit strong non-monotonic behaviour as a function of the chemical potential. 
The non-linear responses we determined involve a term that depends on the in-plane electric and the magnetic fields, and a second-order term in the electric field. For the former term, an anomalous out-of-plane magnetization is generated, exceeding the in-plane EE. 
For the latter term, the out-of-plane magnetization is the only response allowed, and it is enhanced when the chemical potential is close to the crossing of the electronic bands due to RSOC. In both cases, it is remarkable that out-of-plane magnetization is obtained in response only to in-plane fields.
All these features can be exploited to fully characterize the altermagnetic 
nature of the electron gas at the interface.
Indeed, the behaviour of the magnetization with the filling can be directly related to the presence of the spin-orbit, DM interaction and band crossings.
To isolate the essential physics, we focused on a minimal model capturing the local interplay between altermagnetic exchange and Rashba spin-orbit coupling near the $\Gamma$ point. This approach reveals the core mechanism underlying the anomalous Edelstein response in altermagnets. Other band crossings at high-symmetry points can be described within the same framework (see Appendix~\ref{app:tight-binding}), and while a full multiband analysis may introduce quantitative differences, it does not alter the fundamental physical picture.
The uncovered \textit{anomalous} non-linear altermagnetic Edelstein responses may be used experimentally to induce efficient spin-to-charge conversion at altermagnetic interfaces leading to novel electronic and spintronic applications.

\section{Acknowledgments}
We thank Oleg Janson and Carmine Autieri for fruitful discussions. We thank Ulrike Nitzsche for technical assistance. M.T. acknowledges financial support from ``Fondazione Angelo Della Riccia''.
This work was financed by Horizon Europe EIC Pathfinder under the grant IQARO number 101115190. 
R.C. and F.R.  acknowledge funding from Ministero dell’Istruzione, dell’Università e della Ricerca (MIUR) for the PRIN project QUESTIONS (Grant No. PRIN P2022KWFBH) and STIMO (Grant No. PRIN 2022TWZ9NR). This work received support from the PNRR MUR project PE0000023-NQSTI (TOPQIN and SPUNTO). 
This work was supported by the Deutsche Forschungsgemeinschaft (DFG, German Research Foundation) through the Sonderforschungsbereich SFB 1143, Grant No. YE 232/2-1, and under Germany’s Excellence Strategy through the W\"urzburg-Dresden Cluster of Excellence on Complexity and Topology in Quantum Matter – ct.qmat (EXC 2147, Project IDs No. 390858490 and No. 392019).

\appendix

\section{Spin-texture analysis}\label{app:spin_text_an}

In this section we analyze in detail the spin-texture of the Hamiltonian including also the Zeeman coupling 
\begin{equation}
    \hat{H}= \frac{k^2}{2m} + \gamma (k_x^2 - k_y^2)\hat{\sigma}_z + \alpha \hat{z} \cdot (\vec \sigma \times \mathbf{k})+h_y\hat{\sigma}_y-\mu.
\label{eq:ham_altermagnets_rashba_zeeman}
\end{equation}
The eigenvalues of this $2\times 2$ matrix are easily obtained, as well as the direction of the spin quantization axis
\begin{equation}
    \varepsilon^\pm = \frac{k^2}{2m} \pm \sqrt{(h_y - \alpha k_x)^2 + \alpha^2 k_y^2 + \gamma^2(k_x^2 - k_y^2)^2},
\end{equation}
\begin{equation}
    \mathbf{s} = \frac{(\alpha k_y,\, h_y - \alpha k_x,\, \gamma(k_x^2 - k_y^2))}{\sqrt{(h_y - \alpha k_x)^2 + \alpha^2 k_y^2 + \gamma^2(k_x^2 - k_y^2)^2}}.
\end{equation}

\begin{figure}
    \centering
    \includegraphics[trim ={1cm 0 1cm 0}, width=0.4\textwidth]{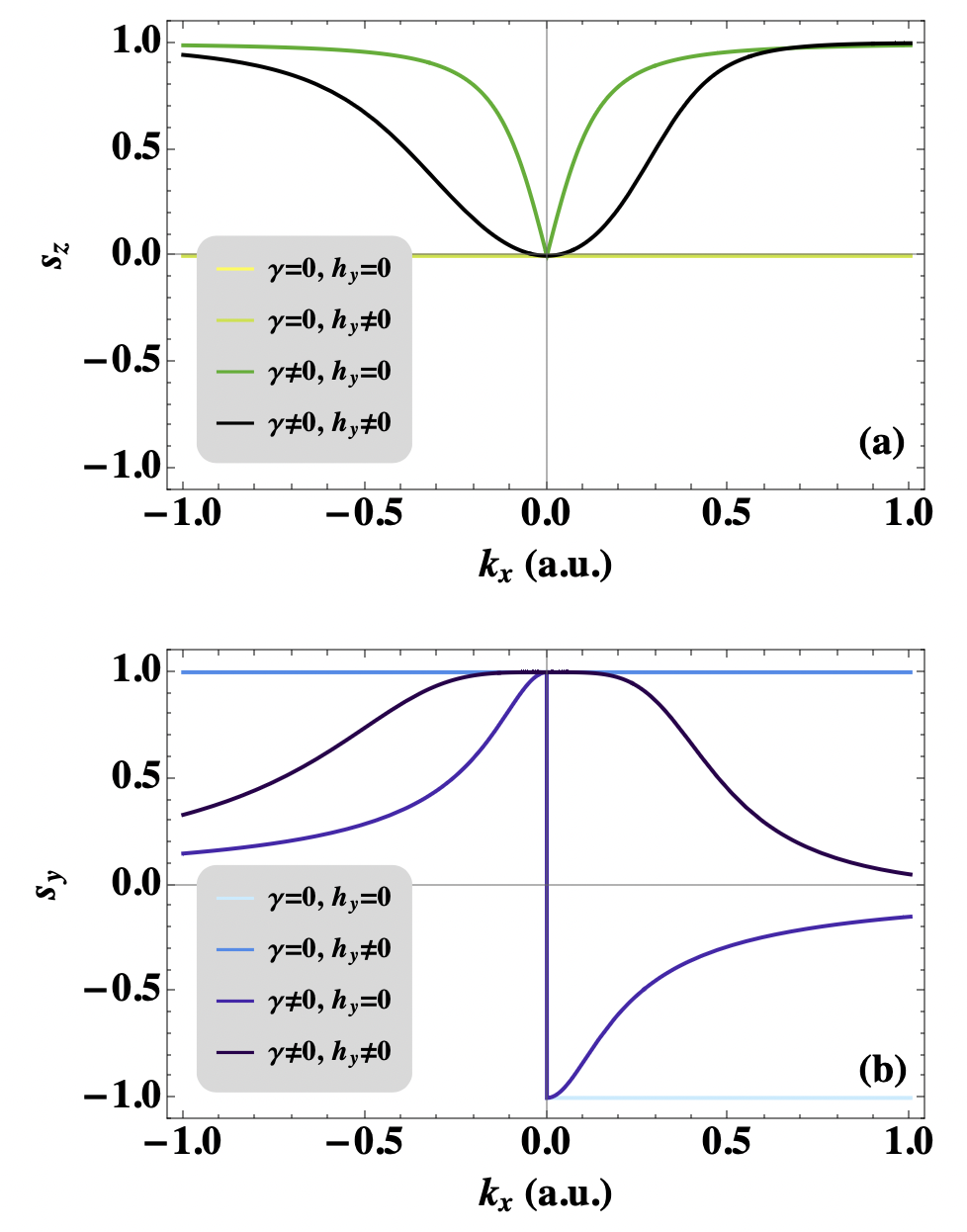}
    \caption{$s_z$ (a) and $s_y$ (b) as a function of $k_x$ and $k_y=0$ with benchmark values of $\alpha\neq0$ and by varying $\gamma$ and $h_y$. The values of $\alpha$, $\gamma$ and $h_y$ have been chosen to make the plot clearer.}
    \label{fig:spin_textures}
\end{figure}

For $ \gamma = 0 $, the band structure exhibits a crossing at $ \mathbf{k} = (h_y/\alpha,\, 0) $. However, the spin texture is tilted along the $ y $-direction, distorting the in-plane chirality. When $ \gamma \neq 0 $, the $ s_z $ component of the spin becomes an odd function of $ \mathbf{k} $, as the parity symmetry is broken by the denominator of $ \mathbf{s} $. \\Notably, for $ \alpha = 0 $, the presence of $ h_y $ alone does not break the even symmetry of $ s_z $ (i.e. $s_z(\mathbf{k})=s_z(-\mathbf{k})$). 
This highlights the need for all three ingredients—Rashba spin-orbit coupling ($ \alpha $), Zeeman field ($ h_y $), and anisotropic coupling ($ \gamma $)—to generate an out-of-plane Edelstein effect.
In Fig.~\ref{fig:spin_textures} we show the values of $s_y$ and $s_z$ as a function of $k_x$ for $k_y=0$ for $\alpha\neq0$ and by varying $\gamma$ and $h_y$.
In case of a pure Rashba coupling only $s_y$ is present and is constant with a change of sign at $k_x=0$, while when a Zeeman field is present along the line $k_y=0$ the spin is constan. When the altermagnetic coupling is turned on, the even (odd) symmetry in $k_x$ is present for $s_z$ ($s_y$), which is broken while $h_y\neq0$.

\section{Ab-initio RuO$_2$ bilayer band structure and parameter fit} \label{app:fit}
\begin{figure}
 \centering
 \includegraphics[width=0.45\textwidth]{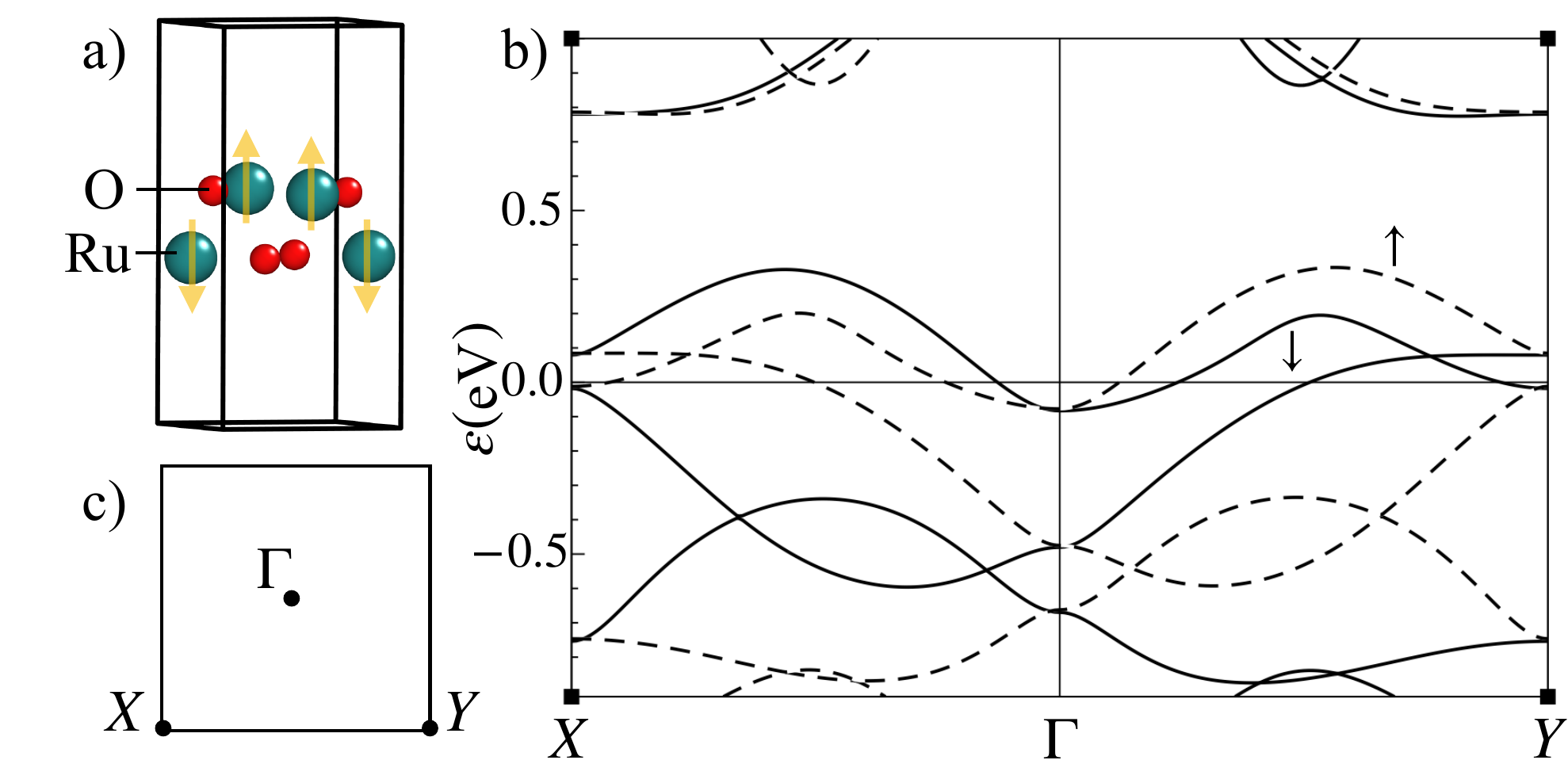}
 \hfill
\includegraphics[width=0.45\textwidth]{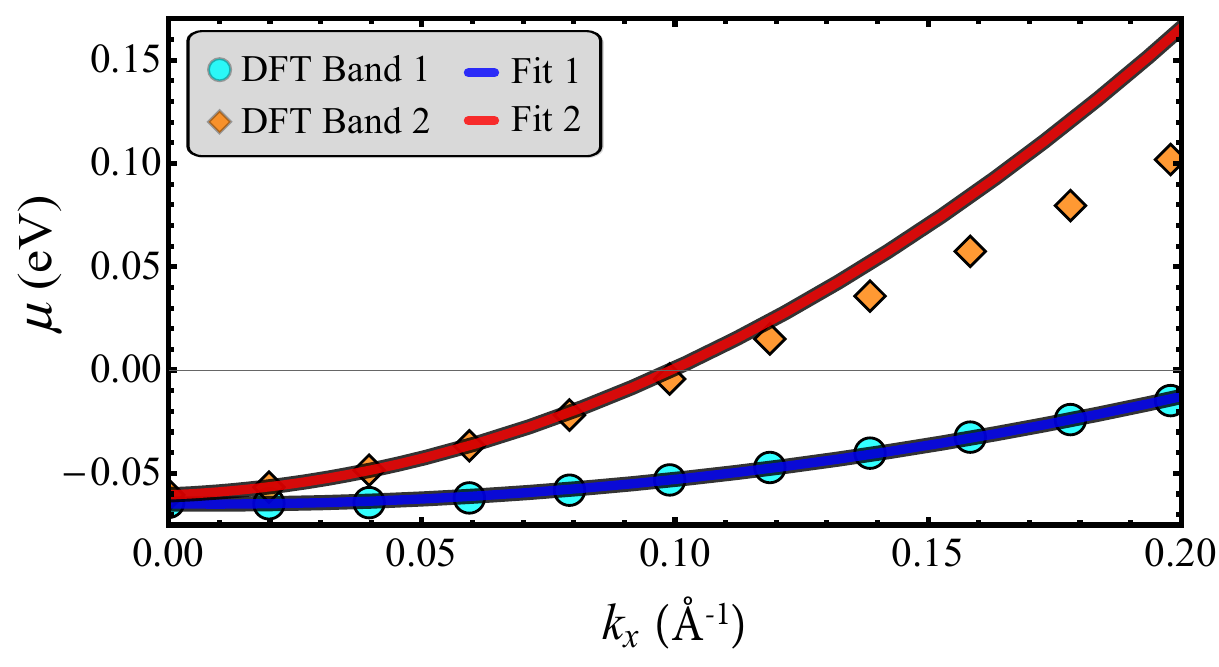}
\begin{picture}(0,0)
   \put(-230,110){ d)}
 \end{picture}
 \caption{(a-c) Bilayer of RuO$_2$ and its electronic structure density-functional calculations. (a) Crystal structure. (b) Band structure along high symmetry points indicated in (c). (d) Fit of the bands crossing the Fermi level nearby $\Gamma$ along the direction $\Gamma Y$.}
 \label{fig:band_strucure}
\end{figure}

Here we estimate the strength of the altermagnetic coupling $\gamma$ by explicit comparison with the local band structure of a RuO$_2$ bilayer.
Based on the bulk crystal structure, we construct a bilayer oriented along the $c$-axis of RuO$_2$ bulk, containing two Ru ions related by fourfold rotoinversion symmetry. The altermagnetic bilayer can be represented with the space group Cmm2 (No. 35), with a unit cell having $a=6.35$~\AA.
For this structure, we perform density-functional calculations in the generalized gradient approximation based on the FPLO code~\cite{kopernik1999full}. We use a $k$-points mesh having $16\times16\times1$ subdivisions along with a tetrahedron method for Brillouin zone integrations. 
Figure~\ref{fig:band_strucure} shows the band structure in the absence of RSOC. We apply our effective two-band model to describe the two bands closest to the Fermi energy at $\Gamma$. 
The parameter $\gamma$ and $m$ of Hamiltonian~\eqref{eq:ham_altermagnets_rashba_zeeman} (for $\alpha=0$ and $h_y=0$) can be extracted from DFT data of RuO$_2$ bilayer. We note here that our model has a spin-ordering rotated of $45^\circ$ with respect to the RuO$_2$, which simplifies the analysis of the Edelstein susceptibility tensor.
In Fig.~\ref{fig:band_strucure} we show the DFT data of RuO$_2$ and the fit of the band structure for the two bands crossed by the Fermi level. The two bands are well fit by
\begin{equation}
 \begin{cases}
 E_-=1.412 \;\text{eV} \text{\AA}^2 \;k_x^2 -0.023 \;\text{eV} \text{\AA}\; k_x -0.065  \;\text{eV}\\
 E_+=5.175\;\text{eV} \text{\AA}^2 \; k_x^2 +0.094\;\text{eV} \text{\AA} \; k_x -0.060 \;\text{eV}.
 \end{cases}
 \label{eq:fit}
\end{equation}
By comparison with the analytical expression of the eigenvalues of Hamiltonian~\eqref{eq:ham_altermagnets_rashba_zeeman}
we obtain $m=0.152$~eV$^{-1}$~\AA$^{-2}$ and $\gamma=1.881$ $\text{eV} \text{\AA}^2$. 

\section{Parametric dependence of the Edelstein response}\label{app:parameters}
\begin{figure}
 \centering
 \includegraphics[width=0.47\textwidth]{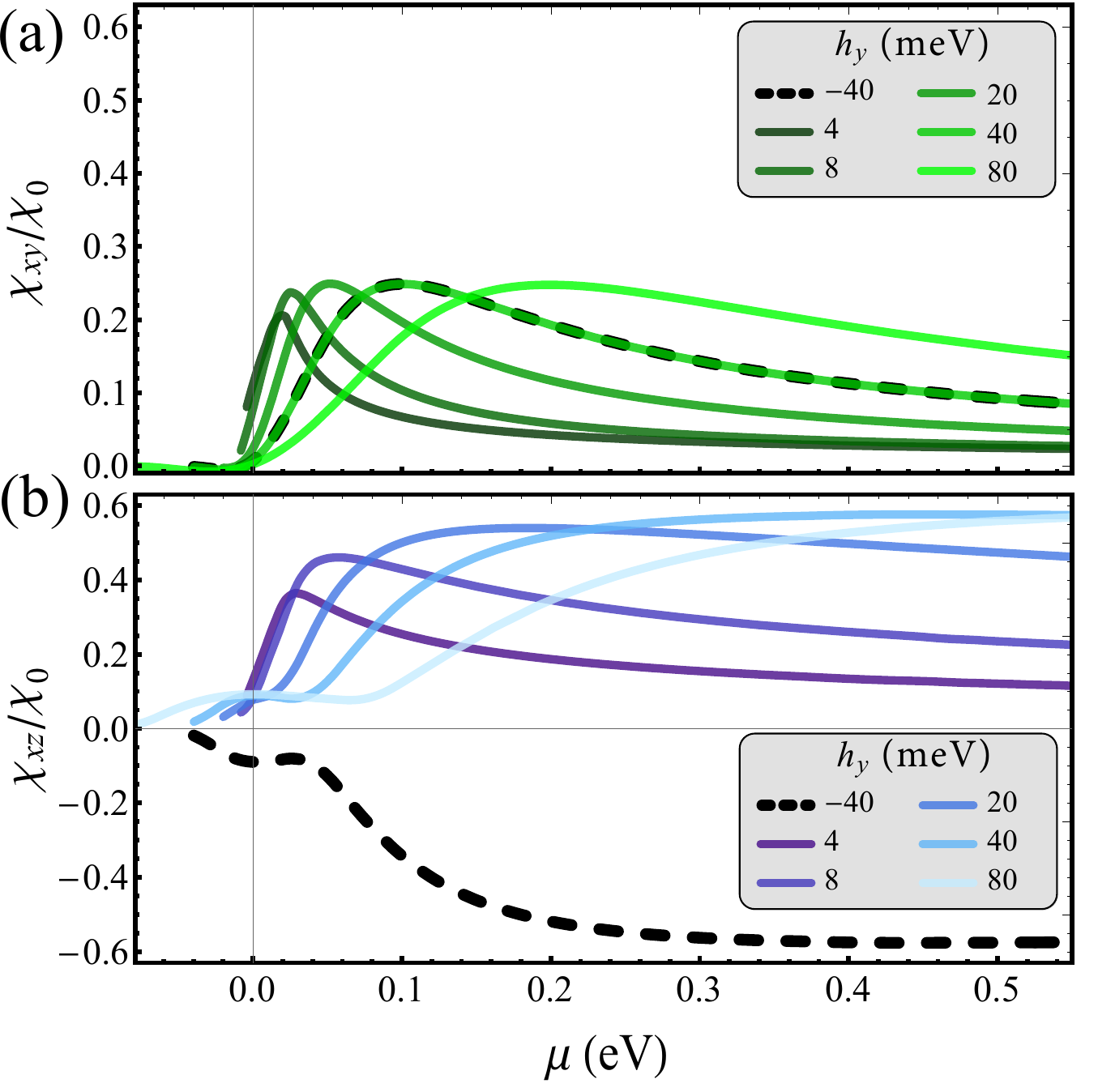}
 \hfill
 \includegraphics[width=0.47\textwidth]{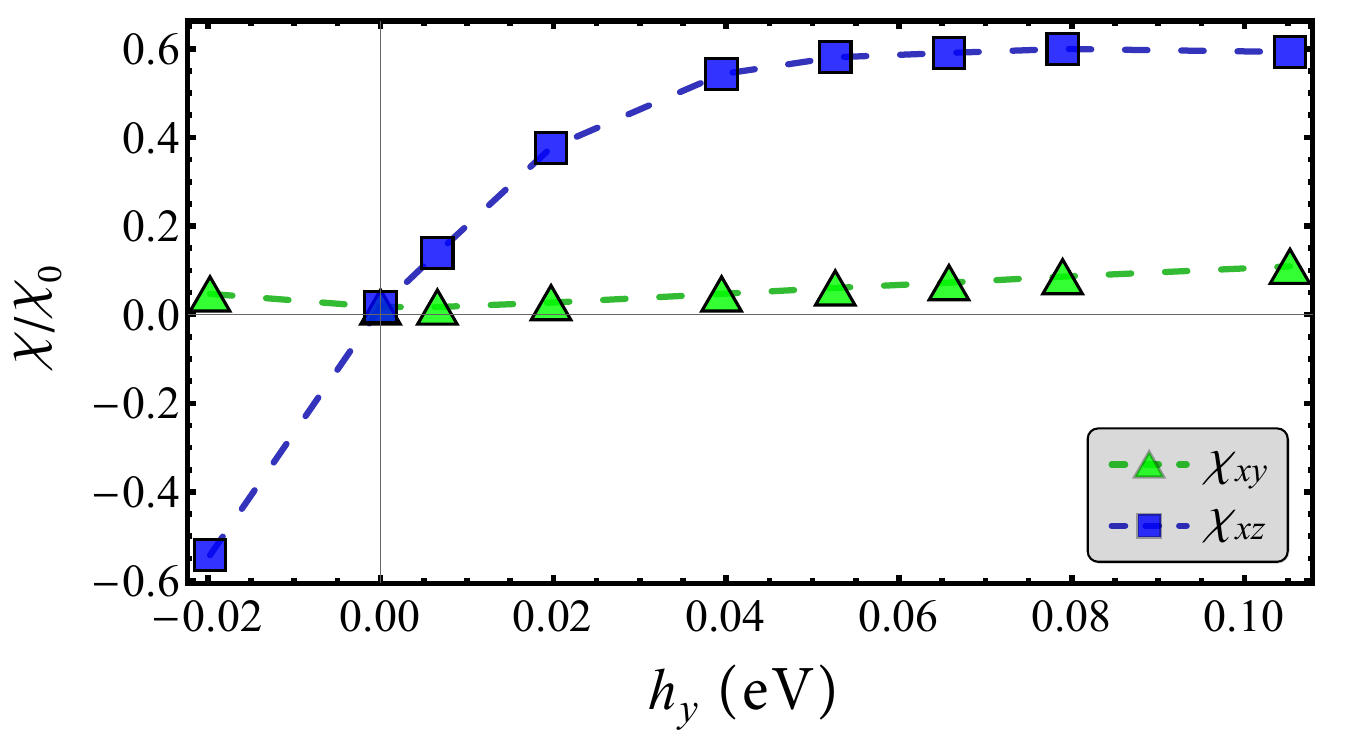}
 \begin{picture}(0,0)
   \put(-240,120){\large \bfseries (c)}
 \end{picture}
 \caption{(a) $\chi_{xy}$ and (b) $\chi_{xz}$ as a function of the chemical potential for $\alpha=52$ meV \AA~and different values of the in-plane Zeeman field $h_y$. (c) Edelstein susceptibility behaviour with Zeeman coupling for $\alpha=52$ meV \AA~and for fixed chemical potential $\mu=1.05$ eV and $\gamma=1.881$~eV~\AA$^{2}$. Lines are guides to the eye.}
 \label{fig:h_beha}
\end{figure}

This section details the dependence of the Edelstein responses on the RSOC parameter ($\alpha$) and the Zeeman coupling ($h_y$). 
In Fig.~\ref{fig:h_beha}, we show the behaviour of the EE for different $h_y$. The in-plane response is even in $h_y$, %remaining identical for $h_y\to -h_y$, 
while the out-of-plane response is odd in $h_y$.
The inversion of the magnetic field $h_y\to -h_y$ can be canceled by a corresponding change in sign $k_x\to -k_x$, which leaves the eigenvalues of the Hamiltonian unchanged. With such a transformation, the group velocity $v_x\to -v_x$, and the same goes for the in-plane spin $s_y\to -s_y$, while the out-of-plane spin, determined by the altermagnetic term, which is even in $k_x$, remains unchanged $s_z\to s_z$. 
Therefore, from the definition of the EE susceptibility, we see that the in-plane response is even, while the out-of-plane response is odd in $h_y$. 
Correspondingly, the out-of-plane response, which vanishes for $h_y\to 0$, must have a linear behavior $\chi_{xz}\propto h_y$ for very small fields.
In Fig.~\ref{fig:h_beha} we show $\chi$ as a function of the Zeeman coupling for a fixed value of the chemical potential $\mu/h_y\gg 1$, in order to show the behvaviour of the $\chi$ in the saturation region.

\begin{figure}
 \centering
 \includegraphics[width=0.47\textwidth]{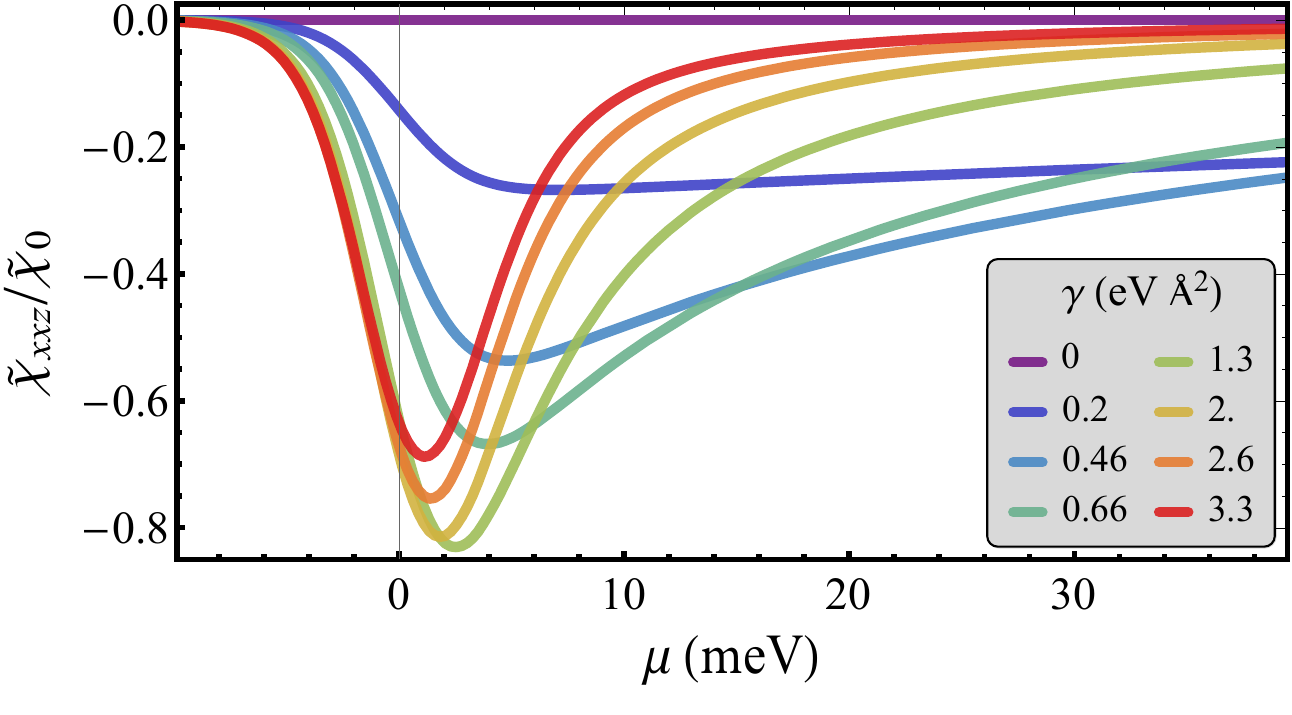}
 \begin{picture}(0,0)
   \put(-240,122){\large \bfseries (a)}
 \end{picture}
 \hfill
 \includegraphics[width=0.47\textwidth]{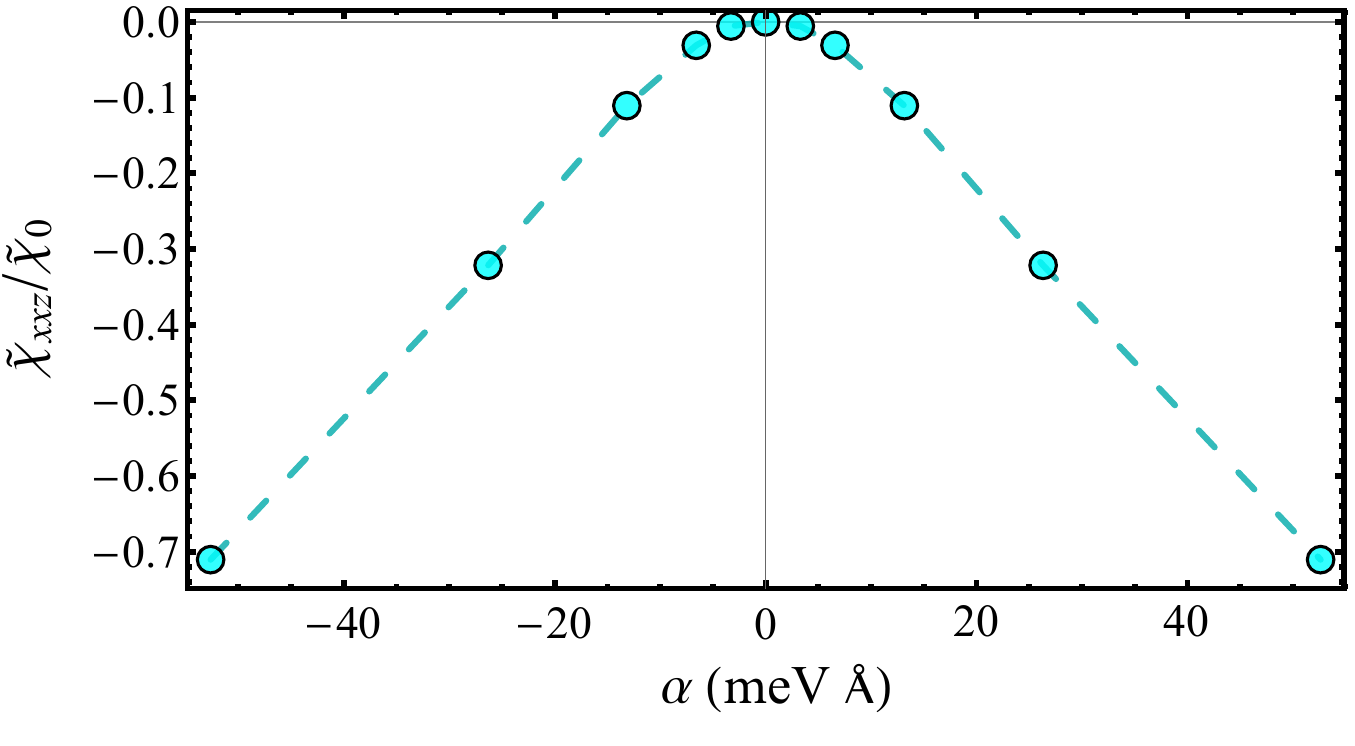}
 \begin{picture}(0,0)
   \put(-240,122){\large \bfseries (b)}
 \end{picture}
 \caption{(a) Second order Edelstein susceptibility as a function of the chemical potential for $\alpha=52$ meV \AA~and different values of $\gamma$. 
 %The dashed vertical line marks the value of $\mu$ used for Fig.~\ref{fig:gamma_dependence}(a). 
 (b) Second-order Edelstein susceptibility $\Tilde{\chi}_{xxz}$ as a function of $\alpha$ for $\mu=0$ and $\gamma=1.881$~eV~\AA$^{2}$. The line is a guide to the eye.}
 \label{fig:gamma_figures}
\end{figure}

\textit{Second-order Edelstein susceptibility---} The ratio between of the RSOC and the altermagnetic energy scales is determined by $|\alpha|/|k_F\gamma|$, with $k_F$ to be the Fermi momentum. It is a measure for their competition, which is strongest when $|\alpha|/|k_F\gamma|\simeq1$. For the value $|\alpha|/|k_F\gamma|\simeq1$ we observe that the second-order anomalous susceptibility has a peak.
In Fig.~\ref{fig:gamma_figures} we show the second-order anomalous Edelstein susceptibility for different values of $\gamma$. 
It tends to peak at a small value of the chemical potential close to the band crossing at $\Gamma$ (see Fig.~\ref{fig:spin_density_rashba} of main text).
The peak is substantially broadened when $|\gamma k_F|>|\alpha|$. Indeed 
the peak-shape of the second-order response is lost for small values of $\gamma$ ($\gamma=0.2$~eV~\AA$^2$). Within the considered range of $\mu$ values, the maximum explored Fermi momentum corresponds to $k_F=0.22$~\AA$^{-1}$, so that $|\alpha|>|\gamma k_F|$. However, since the non-relativistic spin splitting is expected to be larger (as seen from DFT calculation), the peak-shape feature might be used as a probe of the band crossing due to the Rashba coupling. 

Finally, to quantify the effect of the RSOC, we show in Fig.~\ref{fig:gamma_figures} the dependence on $\alpha$ of the second-order Edelstein susceptibility. 
%As expected from Eq.~\eqref{eq:simple_calc}, 
For a fixed chemical potential the susceptibility is an even function of $\alpha$, showing also a sign change for $\alpha$ close to zero.

\section{Boltzmann approach to non-linear EE }\label{app:analytical}

The first- and second-order Edelstein reads
\begin{equation}
    M_k=\chi_{ik}E_i+\Tilde{\chi}_{ijk}E_iE_j,
\end{equation}
where $\chi_{ik}$ and $\Tilde{\chi}_{ijk}$ are the linear and second-order Edelstein susceptibilities, respectively. In the Boltzmann approach, the magnetization is obtained as
\begin{equation}
    M_k=\mu_B\sum_{n\hspace{0.05cm}occ}\int \frac{d^2\mathbf{k}}{(2\pi)^2} s^n_k(\mathbf{k}) f(\mathbf{k}),
\end{equation}
with $s_k^n$ the spin component of band $n$ and $f(\mathbf{k})$ the distribution function.

In the Boltzmann framework, we expand $f=f_{\rm th}+\varphi^{(1)}+\varphi^{(2)}+\cdots$ in powers of $\mathbf{E}$, with the lowest-order $f_{\rm th}$ being the Fermi distribution function, and for homogeneous, stationary transport with constant relaxation time $\tau$, the first- and second-order corrections read
\begin{align}
    \varphi^{(1)} &= |e| \tau\, \mathbf{E} \cdot \nabla_{\mathbf{k}} f_{\rm th}, \\
    \varphi^{(2)} &= e^2 \tau^2\, \frac{\partial}{\partial k_j} \left( \frac{\partial f_{\rm th}}{\partial k_i} \right) E_i E_j.
\end{align}
The second-order susceptibility is then given by
\begin{equation}
    \Tilde{\chi}_{xxz}=-\mu_B e^2 \tau^2 S_{\text{cell}} \sum_{\nu}\int \frac{d^2 \mathbf{k}}{4\pi^2}  \nu \frac{\partial  s_z(\mathbf{k})}{\partial k_x}  \delta(\varepsilon^\nu_{\mathbf{k}}-\mu) v_x^\nu(\mathbf{k}).
    \label{eq:tilde_chi_eq_sup}
\end{equation}
An estimate of the prefactor yields
\begin{equation}
    \Tilde{\chi}_0=\frac{\mu_B e^2 \tau^2}{4\pi^2} S_{\text{cell}} \simeq 2.36 \times 10^{-14} \; \frac{\text{m}^2}{\text{V}^2}.
\end{equation}
We adopt the rigid band approximation, which is adequate for the purpose of our model. In principle, for a more complex multiband structure, corrections to the eigenstates due to a strong (or oscillating) electric field could be considered~\cite{kaplan2024unification,jia2024equivalence}, but these effects are not critical in our approach.
\begin{figure*}[t!!]    \includegraphics[width=0.75\textwidth]{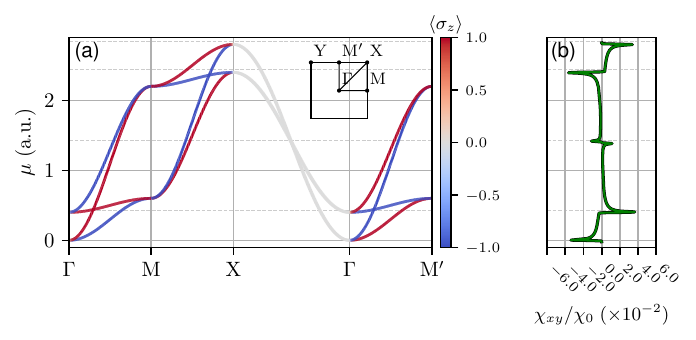}
    \caption{(a) Band structure of the tight-binding model of Hamiltonian~\eqref{eq:TB_model} using: $t=0.6$, $\gamma=0.2$, $t_1=0.05$, and $\alpha=0.01$. The energy has been shifted in order to set the minimum of the bands to $\mu=0$. (b) Edelstein susceptibility $\chi_{xy}/\chi_{0}$ as a function of the chemical potential.}
    \label{fig:Tb_bands}
\end{figure*}
In the absence of Rashba coupling it is straightforward to see that $\frac{\partial s_z}{\partial k_x}=0$ and therefore $\chi_{xxz}=0$. We use these expressions to determine the non-linear Edelstein response shown in the main text.

\section{Tight-binding model}\label{app:tight-binding}
In this section we highlight the role of the crossing at high-symmetry points and the multi-band crossings on the linear Edelstein response. 
We construct a minimal tight-binding model from the irreducible representations of the altermagnetic symmetry group adopted in the main text, including the inversion-breaking Rashba term
\begin{dmath}
    H(\mathbf{k}) = 
    [- 2t (\cos k_x + \cos k_y)\,  -\mu]\sigma_0 \tau_0
    -2\gamma (\cos k_x - \cos k_y)\, \sigma_z \tau_x \nonumber 
     -2t_1 (\cos k_x + \cos k_y)\,\sigma_0  \tau_x  
+ \alpha \left[ \sin k_x\,  \sigma_y - \sin k_y\,  \sigma_x \right]\tau_x.
    \label{eq:TB_model}
\end{dmath}
Here $t_1$ has been included to split the four-fold degeneracy. For $t_1=0$ and $t=1/2m$ the 
%Using $t=1/(2m)$, 
the expansion at ${\mathbf{k}}\approx0$ is equivalent to Ham.~\eqref{eq:ham_altermagnets_rashba_zeeman}. The band structure is shown in Fig.~\ref{fig:Tb_bands}, together with the Edelstein response function for varying filling of the bands.
The dominant contribution to the Edelstein response comes from the band crossings, as one could expect. In particular, the band crossings at the $\Gamma$ points, which give the largest contributions, are well described by our $\bf k\cdot p$ local models, providing a concrete example of the local nature of the Edelstein response. At large fillings, a specular contribution comes from the band crossings at X and Y, due to a change of chirality of the spin texture close to the corner of the Brillouin zone. Band crossings at M and M' do not produce an Edelstein response, since the Rashba operator has vanishing matrix elements between the doubly degenerate eigenstates. This is immediately visible by expanding the Hamiltonian~\eqref{eq:TB_model}
$\mathbf{q}_{\Delta\mathbf{k}}={\bf k}-\Delta{\bf k}$ with $\Delta{\bf k}=(\pm \pi,0)$
\begin{dmath}
    H(\mathbf{q}_{(\pi,0)})\approx t (-q_x^2+q_y^2)\sigma_0\tau_0 +\gamma (4 -q_x^2-q_y^2)\sigma_z\tau_x +t_1 (-q_x^2+q_y^2)\sigma_0\tau_x +\alpha(-q_x \sigma_y-q_y\sigma_x)\tau_x,
\end{dmath}
so for $\mathbf{q}\to 0$ the doubly degenerate eigenstates with $\sigma_z \tau_x=+1$ (or $\sigma_z \tau_x=-1$) lead to a zero matrix element for the RSOC.

%and for $q\to 0$ leaves $4\gamma \sigma_z\tau_x$ term which dominates.
%
Another band crossing occurs at $k_x=k_y=\pi/2$; a local expansion gives 
\begin{dmath}
    H(\mathbf{q}_{(\pi/2,\pi/2)})\approx 2t (q_x+q_y)\sigma_0\tau_0 +2\gamma (q_x-q_y)\sigma_z\tau_x +2t_1 (q_x+q_y)\sigma_0\tau_x +\alpha((1-q_x^2) \sigma_y+(-1+q_y^2)\sigma_x)\tau_x.
\end{dmath}
For $q\to0$, the Rashba term reduces to $\alpha(\sigma_y -\sigma_x)\tau_x$, so the crossing is gapped by the RSOC which enforces an in-plane spin. At finite $\bf q$, the spin tilts out of plane due to the altermagnetic terms, enforcing a spin-momentum locking and inducing a symmetry-allowed Edelstein response, which is clearly visible at half-filling, albeit of lower intensity than the response coming from the $\Gamma$ and X points.

\section{Intrinsic interband contribution}\label{app:intrinsic}
In this section, we discuss the role of the interband intrinsic contribution to the Edelstein effect in our model. Its general definition~\cite{li2015intraband} is    

\begin{equation}
    \chi^{\text{int}}_{ij} = e\hbar
    \int \frac{d^2 {\bf k}}{(2\pi)^2} \sum_{m\neq n}
    \textrm{Im}\left[\frac{s_j^{nm}(\mathbf k) v_i^{mn}(\mathbf k)}{(\varepsilon_{\bf k}^n-\varepsilon_{\bf k}^m)^2}\right]
    \bigl[f(\varepsilon_{\bf k}^n)-f(\varepsilon_{\bf k}^m)\bigr],
    \label{eq:intrs-app}
\end{equation}
where $s_j^{nm}({\bf k})=\mel{u_n({\bf k})}{\sigma_j}{u_m({\bf k})}$ and $v_i^{mn}({\bf k})=\mel{u_m({\bf k})}{\partial_{k_j}H({\bf k})}{u_n({\bf k})}$ are the off-difagonal matrix elements of the Pauli matrix and group velocity operators over the eigenstates of the Hamiltonian.

This contribution has to be compared with the extrinsic Edelstein susceptibility  

\begin{equation}
    \chi^{\mathrm{ext}}_{ij}
    = e \tau \sum_{n}
    \int_{\mathrm{BZ}} \frac{d^2 {\bf k}}{(2\pi)^2}\;
    \Big(-\frac{\partial f_0}{\partial \varepsilon^n_{\mathbf k}}\Big)\,
    v^{n}_{i}(\mathbf k)\,
    s^{n}_{j}(\mathbf k),
    \label{eq:ext-app}
\end{equation}
where all quantities are expressed in standard units.  
From these definitions, it becomes immediately clear that the intrinsic contribution can dominate over the extrinsic one only over those regions of the Brillouin zone where $\varepsilon_m-\varepsilon_n\lesssim \hbar/\tau$ (excepting cases where the extrinsic contribution vanishes due to some symmetry, which we have established not to be the case for the components of the Edelstein susceptibility we determine). Since $\hbar/\tau\sim 0.7\,\mathrm{meV}$, whereas all the energy scales within our model Hamiltonian are significantly larger, clearly the interband intrinsic contribution can become competitive only close to near-degeneracy points. Let us therefore consider such contribution.
If Zeeman terms are included, the spectrum is fully gapped, with typical gaps larger than $\sim 50\,\mathrm{meV}$. In this regime, the intrinsic susceptibility is always negligible compared to the extrinsic one.  
The only case requiring more careful analysis is the gapless limit at ${\bf k}=0$. By considering Hamiltonian~\eqref{eq:ham_altermagnets_rashba}, and using Eq.~\eqref{eq:intrs-app}, one obtains for the integrand of $\chi_{ix}=e\hbar\int I(s_i)\, d^2{\bf k}$ 
\begin{equation}
    I(s_z)= -\frac{1}{4\pi^2}\frac{k_y \alpha^{2}}{4 \left[ \bigl( k_x^{2} + k_y^{2} \bigr) \alpha^{2} + \bigl( k_x^{2} - k_y^{2} \bigr)^{2} \gamma^{2} \right]^{3/2}},
\end{equation}
and  
\begin{equation}
    I(s_y)=k_x k_y \,\mathcal{F}(k_x^2,k_y^2),
\end{equation}
where $\mathcal{F}(k_x^2,k_y^2)$ is an even function of ${\bf k}$. Both integrands vanish upon Brillouin-zone integration by symmetry, so the intrinsic contribution is absent in this two-band model.   

\bibliographystyle{unsrt}
\bibliography{edelstein-altermagnetism}

\end{document}